\documentclass[12pt]{article}
\usepackage{amsfonts}
\usepackage{fullpage}
\usepackage{hyperref}
\usepackage{amsmath}
\usepackage{amssymb}
\usepackage{latexsym}
\usepackage{amsmath,amssymb,amsthm}
\usepackage{epsfig}
\usepackage[right=0.8in, top=1in, bottom=1.2in, left=0.8in]{geometry}
\usepackage{setspace}
\usepackage{bm}

\spacing{1.06}
\newtheorem{theorem}{Theorem}

\newtheorem{proposition}{Proposition}

\theoremstyle{remark}

\input{tcilatex}

\begin{document}

\title{Malliavin-based Multilevel Monte Carlo Estimators for Densities of
Max-stable Processes}
\author{Blanchet, J. \thanks{Support from NSF grant DMS-132055 and NSF grant CMMI-1538217 is gratefully acknowledged by J. Blanchet} \and Liu, Z.}
\maketitle

\begin{abstract}
We introduce a class of unbiased Monte Carlo estimators for the multivariate
density of max-stable fields generated by Gaussian processes. Our estimators
take advantage of recent results on exact simulation of max-stable fields
combined with identities studied in the Malliavin calculus literature and
ideas developed in the multilevel Monte Carlo literature. Our approach
allows estimating multivariate densities of max-stable fields with precision 
$\varepsilon $ at a computational cost of order $O\left( \varepsilon
^{-2}\log \log \log \left( 1/\varepsilon \right) \right) $.
\end{abstract}

\section{Introduction}

Max-stable random fields arise as the asymptotic limit of suitably
normalized maxima of many i.i.d. random fields. Intuitively, max-stable
fields are utilized to study the extreme behavior of spatial statistics. For
instance, if the logarithm of a precipitation field during a relatively
short time span follows a Gaussian random field, then extreme precipitations
over a long time horizon, which are obtained by taking the maximum at each
location of many precipitation fields can be argued (if enough temporal
independence can be assumed) to follow a suitable max-stable process.
Precisely these types of applications in environmental science motivate the
study of max-stable processes (see, for example, \cite{EKMBK} for a recent
study of this type).

To calibrate and estimate max-stable random fields, it is desirable to
evaluate the density over a finite set of locations (i.e. multivariate
density of finite-dimensional coordinates of the max-stable field). As we
shall explain, this task becomes prohibitively difficult as the number of
locations increases. This is precisely the motivation behind our
contribution in this paper, which we shall explain more precisely, but
first, we must introduce some basic facts about max-stable processes.

We will focus on a class of max-stable random fields which are driven by
Gaussian processes. These max-stable fields are popular in practice because
their spatial dependence structure is inherited from the underlying Gaussian
covariance structure.

To introduce the max-stable field of interest, let us first fix its domain $%
T\subseteq \mathbb{R}^{m}$, for $m\geq 1$. We introduce a sequence, $\left(
X_{n}\left( \cdot \right) \right) $, of independent and identically
distributed\ copies of a centered Gaussian random field, $X\left( \cdot
\right) =\left( X\left( t\right) :t\in T\right) $. We let $\left(
A_{n}\right) $ be the sequence of arrivals in a Poisson process with unit
rate and independent of $\left( X_{n}\left( \cdot \right) \right) $.

Finally, given a deterministic and bounded function, $\mu :T\longrightarrow 
\mathbb{R}$, we will focus on developing Monte Carlo methods for the finite
dimensional densities of the max-stable field%
\begin{equation}
M(t)=\sup_{n\geq 1}\big\{-\log A_{n}+X_{n}(t)+\mu (t)\big\},\qquad t\in T\,.
\label{eq_defM}
\end{equation}%
(The name max-stable is justified because $M\left( \cdot \right) $ turns out
to satisfy a distributional equation involving the maximum of i.i.d.
centered and normalized copies of $M\left( \cdot \right) $.)

An elegant argument involving Poisson point processes (see \cite{Rib2013})
allows us to conclude that%
\begin{eqnarray}
&&P\left( M\left( t_{1}\right) \leq x_{1},...,M\left( t_{d}\right) \leq
x_{d}\right)   \label{ID_JOIN} \\
&=&\exp \left( E\left[ \max_{i=1}^{d}\{\exp \left( X\left( t_{i}\right) +\mu
\left( t_{i}\right) -x_{i}\right) \}\right] \right) .  \notag
\end{eqnarray}

By redefining $x_{i}$ as $x_{i}-\mu \left( t_{i}\right) $, we might assume
without loss of generality, for the purpose of computing the density of $%
M=\left( M\left( t_{1}\right) ,...,M\left( t_{d}\right) \right) ^{T}$, that $%
\mu \left( t_{i}\right) =0$. We will keep imposing this assumption
throughout the rest of the paper.

Throughout the paper we will keep the number of locations, $d$, over which $%
M\left( \cdot \right) $ is observed, fixed. So, $M$ will remain a $d$%
-dimensional vector throughout our discussion. To avoid confusion between $M$
and $M(.)$, note that we use $M(\cdot )$ when discussing the whole
max-stable field. We will maintain this convention throughout the rest of
the paper for the field $M(\cdot )$ as well as the fields $X_{n}\left( \cdot
\right) $, $n \ge 1$.

The joint density of $M$ can be obtained by subsequent differentiation of (%
\ref{ID_JOIN}) with respect to $x_{1},...,x_{d}$. However, the final
expression obtained for the density contains exponentially many terms. So,
computing the density of $M$ using this direct approach becomes quickly
intractable, even for moderate values of $d$. For example, \cite{Rib2013}
argues that even for $d=10$ one obtains a sum of more than $10^{5}$ terms.

We will construct an unbiased estimator for the density, $f\left( x\right) $%
, of $M$ evaluated at $x=\left( x_{1},...,x_{d}\right) $ for $d\geq 3$. The
construction of our estimator, denoted as $V(x)$, is explained in Section %
\ref{Section_Estimator}. Implementing our estimator avoids the exponential
growth issues which arise if one attempts to evaluate the density directly.
We concentrate on $d\geq 3$ because the case $d=2$ leads to only four terms
which can be easily computed as explained in \cite{HD2013}. More precisely
our contributions are as follows:

\begin{enumerate}
%\item We construct an estimator, $V\left( x\right) $, which is explained in
%Section \ref{Section_Estimator}.

\item The properties of $V\left( x\right) $ are summarized in Section \ref%
{Sect_Main}. In particular, $f\left( x\right) =E\left( V\left( x\right)
\right) $, $Var\left( V\left( x\right) \right) <\infty $, and given a
computational budget of size $b$, we provide a limit theorem which can be
used to estimate $f\left( x\right) $ with complexity $O\left( \left( b\cdot
\log \log \log \left( b\right) \right) ^{2}\right) $ for an error of order $%
O\left( 1/b\right) $ -- see Theorem \ref{Thm_Main} and its discussion.

\item As far as we know this is the first estimator which uses Malliavin
calculus in the context of max-stable density estimation. We believe that
the techniques that we introduce are of independent interest in other areas
in which Malliavin calculus has been used to construct Monte Carlo
estimators. For example, we highlight the following contributions in this
regard,

\begin{enumerate}
\item We introduce a technique which can be used to estimate the density of
the (coordinate-wise) maximum of multivariate variables. We apply this
technique to the case of independent Gaussian vectors, but the technique can
be used more generally, see the development in Section \ref{Sec_Finite_Max}.

\item We explain how to extend the technique in item 3.a) to the case of the
maxima of infinitely many variables. This extension, which is explained in
Section \ref{Sec_Infinite_Max}, highlights the role of a recently introduced
record-breaking technique for the exact sampling of variables such as $M$.

\item We introduce a perturbation technique which controls the variance of
so-called Malliavin-Thalmaier estimators (which are explained in Section \ref%
{Sec_Malliavin_Intro}). These types of estimators have been used to compute
densities of multivariate diffusions (see \cite{KY2009}). Our perturbation
technique, introduced in Section \ref{Sec_Mall_Var}, can be directly used to
improve upon the density estimators in \cite{KY2009}, enabling a
close-to-optimal Monte Carlo rate of convergence for density estimation of
multivariate diffusions.
\end{enumerate}

\item The perturbation technique in Section \ref{Sec_Mall_Var} is combined
with randomized multilevel Monte techniques (see \cite{RG2015} and \cite%
{McL2011}) in order to achieve the following. Starting from an infinite
variance estimator, we introduce a perturbation which makes the estimator
biased, but with finite variance. The randomized multilevel Monte Carlo
technique is then used to remove the bias while keeping the variance finite.
The price to pay is a small degradation in the rate of convergence in the
associated Central Limit Theorem for confidence interval estimation. Instead
of an error rate of order $O(1/b^{1/2})$ as a function of the computational
budget $b$, which is the typical rate, we obtain a rate of order $O\left(
\left( \log \log \log \left( b\right) \right) ^{1/2}/b^{1/2}\right) $. The
Central Limit Theorem is obtained using recently developed results in \cite%
{ZBG2017}.
\end{enumerate}

The rest of the paper is organized as follows. In Section \ref%
{Section_Gen_Strategy} we explain step-by-step, at a high level, the
construction of our estimator. The final form of our estimator is given in
Section \ref{Section_Estimator}. The properties of our estimator are
summarized in Section \ref{Sect_Main}. A numerical experiment is given in
Section \ref{Section_Numerics}. Finally, the details of the implementation
of our estimator, in the form of pseudo-codes, are given in the appendix,
namely, Section \ref{Sec_App}.

\section{General Strategy and Background\label{Section_Gen_Strategy}}

The general strategy is explained in several steps. We first review the
Malliavin-Thalmaier identity by providing a brief explanation of its origins
and connections to classical potential theory. We finish the first step by
noting that there are several disadvantages of the identity, having to do
with variance properties of the estimator and the implicit assumption that a
great degree of information is assumed about the density which we want to
estimate. The subsequent steps in our construction are designed to address
these disadvantages.

In the second step of our construction, we introduce a series of
manipulations which enable the application of the Malliavin-Thalmaier
indirectly, by working only with the $X_{n}$s. These manipulations are
performed assuming that only finitely many Gaussian elements are considered
in the description of $M$.

The third step deals with the fact that the description of $M$ contains
infinitely many Gaussian elements. So, first, we need to explain how to
sample $M$ exactly. We utilize a recently developed algorithm by \cite%
{LBDM2016}. Based on this algorithm, we explain how to extend the
construction from the second step in order to obtain a direct
Malliavin-Thalmaier estimator for the density of $M$.

The fourth step of our construction deals with the fact that a direct
Malliavin-Thalmaier estimator will generally have infinite variance. We
introduce a small random perturbation to remove the singularity appearing in
the Malliavin-Thalmaier estimator, which is the source of the poor variance
performance. Unfortunately, such perturbation also introduces bias in the
estimator.

In order to remove the bias we then apply randomized multilevel Monte Carlo
(see \cite{RG2015} and \cite{McL2011}). Our resulting estimator then is
unbiased and has finite variance as we explain in Section \ref{Sect_Main}.
The price to pay is a small degradation in the rate of convergence of the
associated Central Limit Theorem to obtain confidence intervals.

\subsection{Step 1: The Malliavin-Thalmaier Identity\label%
{Sec_Malliavin_Intro}}

The initial idea behind the construction of our estimator comes from the
Malliavin-Thalmaier identity, which we shall briefly explain. First, recall
the Newtonian potential, given by 
\begin{equation*}
G\left( x\right) =\kappa _{d}\frac{1}{\left\Vert x\right\Vert _{2}^{d-2}},
\end{equation*}%
with $\kappa _{d}=\left( d\left( 2-d\right) \omega _{d}\right) ^{-1}$, where 
$\omega _{d}$ is the volume of a unit ball in $d$ dimensions, for $d\geq 3$.
It is well known, see \cite{Evans2010}, that $G(\cdot)$ satisfies the
equation 
\begin{equation*}
\Delta G\left( x-y\right) =\delta \left( x-y\right)
\end{equation*}%
in the sense of distributions (where $\delta \left( x\right) $ is the delta
function). Therefore, if $M\in R^{d}$ has density $f\left( \mathbf{\cdot }%
\right) $ we can write 
\begin{equation}
f\left( x\right) =\int f\left( y\right) \Delta G\left( x-y\right) dy=E\left(
\Delta G\left( x-M\right) \right) .  \label{PoE}
\end{equation}%
But the previous identity cannot be implemented directly because $G\left(
\cdot \right) $ is harmonic, that is, one can easily verify that $\Delta
G\left( x\right) =0$ for $x\neq 0$ (which is not surprising given that one
expects $\Delta G$ to act as a delta function). The key insight of Malliavin
and Thalmaier is to apply integration by parts in the expression (\ref{PoE}%
). So, let us define%
\begin{equation*}
G_{i}\left( x\right) =\frac{\partial G\left( x\right) }{\partial x_{i}}%
=\left( 2-d\right) \kappa _{d}\frac{x_{i}}{\left\Vert x\right\Vert _{2}^{d}},
\end{equation*}%
and therefore write 
\begin{equation*}
\Delta G\left( x-y\right) =\sum_{i=1}^{d}\frac{\partial ^{2}G\left(
x-y\right) }{\partial x_{i}^{2}}=\sum_{i=1}^{d}\frac{\partial G_{i}\left(
x-y\right) }{\partial x_{i}}.
\end{equation*}%
Consequently, because 
\begin{equation*}
\frac{\partial G_{i}\left( x-y\right) }{\partial x_{i}}=-\frac{\partial
G_{i}\left( x-y\right) }{\partial y_{i}},
\end{equation*}%
we have that 
\begin{eqnarray*}
&&\int ...\int \frac{\partial G_{i}\left( x-y\right) }{\partial x_{i}}%
f\left( y_{1},...,y_{d}\right) dy_{1}dy_{2}....dy_{d} \\
&=&-\int ...\int \frac{\partial G_{i}\left( x-y\right) }{\partial y_{i}}%
f\left( y_{1},...,y_{d}\right) dy_{1}dy_{2}....dy_{d} \\
&=&\int ...\int G_{i}\left( x-y\right) \frac{\partial f\left(
y_{1},...,y_{d}\right) }{\partial y_{i}}dy_{1}dy_{2}....dy_{d} \\
&=&E\left( G_{i}\left( x-M\right) \frac{\partial }{\partial y_{i}}\log
f\left( M\right) \right) .
\end{eqnarray*}%
Therefore, we arrive at the following Malliavin-Thalmaier%
\begin{equation}
f\left( x\right) =\sum_{i=1}^{d}E\left( G_{i}\left( x-M\right) \frac{%
\partial }{\partial y_{i}}\log f\left( M\right) \right) .  \label{Bas_MT_I}
\end{equation}%
Refer to \cite{MT2006} and \cite{KY2009} for rigorous proof of this identity.

There are two immediate concerns when applying the Malliavin-Thalmaier
identity. First, a direct use of the identity requires some basic knowledge
of the density of interest, which is precisely the quantity that we wish to
estimate. The second issue, which is not evident from (\ref{Bas_MT_I}), is
that the singularity which arises when $x=M$ in the definition of $%
G_{i}\left( x-M\right) $, causes the estimator (\ref{Bas_MT_I}) to typically
have infinite variance.

\subsection{Step 2: Applying the Malliavin-Thalmaier Identity to Finite
Maxima \label{Sec_Finite_Max}}

We now shall explain how to address the first issue discussed at the end of
the previous subsection. Define%
\begin{equation*}
M_{n}\left( t\right) =\max_{k=1}^{n}\{-\log \left( A_{k}\right) +X_{k}\left(
t\right) \},
\end{equation*}%
and put $M_{n}=\left( M_{n}\left( t_{1}\right) ,...M_{n}\left( t_{d}\right)
\right) ^{T}$. Note that%
\begin{equation}
\frac{\partial G_{i}\left( x-M_{n}\right) }{\partial x_{i}}=-\frac{\partial
G_{i}\left( x-M_{n}\right) }{\partial M_{n}\left( t_{i}\right) }.
\label{Eq_D0}
\end{equation}%
In turn, by the chain rule, 
\begin{equation}
\frac{\partial G_{i}\left( x-M_{n}\right) }{\partial X_{k}\left(
t_{i}\right) }=\frac{\partial G_{i}\left( x-M_{n}\right) }{\partial
M_{n}\left( t_{i}\right) }\frac{\partial M_{n}\left( t_{i}\right) }{\partial
X_{k}\left( t_{i}\right) }.  \label{Eq_D1}
\end{equation}%
Further, with probability one (due to the fact that $(A_1, A_2, \ldots,
A_{k})$ has a density),%
\begin{equation*}
\sum_{k=1}^{n}\frac{\partial M_{n}\left( t_{i}\right) }{\partial X_{k}\left(
t_{i}\right) }=\sum_{k=1}^{n}I\left( M_{n}\left( t_{i}\right) =X_{k}\left(
t_{i}\right) -\log \left( A_{k}\right) \right) =1.
\end{equation*}%
Consequently, from equation (\ref{Eq_D1}) we conclude that%
\begin{equation*}
\sum_{k=1}^{n}\frac{\partial G_{i}\left( x-M_{n}\right) }{\partial
X_{k}\left( t_{i}\right) }=\frac{\partial G_{i}\left( x-M_{n}\right) }{%
\partial M_{n}\left( t_{i}\right) },
\end{equation*}%
and therefore, from (\ref{Eq_D0}), we obtain $\ $

\begin{equation*}
\frac{\partial G_{i}\left( x-M_{n}\right) }{\partial x_{i}}=-\sum_{k=1}^{n}%
\frac{\partial G_{i}\left( x-M_{n}\right) }{\partial X_{k}\left(
t_{i}\right) }.
\end{equation*}

We now can apply integration by parts as we did in our derivation of (\ref%
{Bas_MT_I}). The difference is that the density of $X_{k}=\left( X_{k}\left(
t_{1}\right) ,...,X_{k}\left( t_{d}\right) \right) ^{T}$ is known and
therefore we obtain that%
\begin{equation*}
E\left( \frac{\partial G_{i}\left( x-M_{n}\right) }{\partial X_{k}\left(
t_{i}\right) }\right) =E\left( G_{i}\left( x-M_{n}\right) \cdot
e_{i}^{T}\Sigma ^{-1}X_{k}\right) ,
\end{equation*}%
where $e_{i}$ is the $i$-th vector in the canonical basis in Euclidean space.

Consequently, we conclude that 
\begin{eqnarray*}
E\left( \frac{\partial G_{i}\left( x-M_{n}\right) }{\partial x_{i}}\right)
&=&-\sum_{k=1}^{n}E\left( \frac{\partial G_{i}\left( x-M_{n}\right) }{%
\partial X_{k}\left( t_{i}\right) }\right) \\
&=&-E\left( G_{i}\left( x-M_{n}\right) \cdot e_{i}^{T}\Sigma
^{-1}\sum_{k=1}^{n}X_{k}\right) .
\end{eqnarray*}

In summary, if $f_{n}$ is the density of $M_{n}$ we have that 
\begin{eqnarray}
f_{n}\left( x_{1},...,x_{d}\right) &=&E\left( \sum_{i=1}^{d}\frac{\partial
G_{i}\left( x-M_{n}\right) }{\partial x_{i}}\right)  \label{Est_fn} \\
&=&-E\left( \sum_{i=1}^{d}\sum_{k=1}^{n}G_{i}\left( x-M_{n}\right) \cdot
e_{i}^{T}\Sigma ^{-1}X_{k}\right) .  \notag
\end{eqnarray}%
The verification of this identity follows a very similar argument as that
provided for the proof of (\ref{Bas_MT_I}) in \cite{MT2006}.

\subsection{Step 3: Extending the Malliavin-Thalmaier Identity to Infinite
Maxima\label{Sec_Infinite_Max}}

In order to extend the definition of the estimator (\ref{Est_fn}), we wish
to send $n\rightarrow \infty $ and obtain a simulatable expression of an
estimator. Because we will be using a recently developed estimator for $M$
in \cite{LBDM2016}, we need to impose the following assumptions on $%
X_{n}\left( \cdot \right) $.

\begin{enumerate}
\item[B1)] In addition to assuming $E[X_{n}(t)]=0$, we write $\sigma
^{2}(t)=Var\left( X\left( t\right) \right) $.

\item[B2)] Assume that $\bar{\sigma}=\sup_{t\in T}\sigma(t)<\infty $ and $%
\sup_{t\in T}\left\vert \mu \left( t\right) \right\vert <\infty $.

\item[B3)] Suppose that $E\exp \left( \sup_{t\in T}X\left( t\right) \right)
<\infty $.
\end{enumerate}

A key element of the algorithm in \cite{LBDM2016} is the idea of record
breakers. In order to describe this idea, let us write $\left\Vert
X_{n}\right\Vert _{\infty }=\max_{i=1,\ldots ,d}\left\vert
X_{n}(t_{i})\right\vert $.

Following the development in \cite{LBDM2016} we can identify three random
times as follows.

The first is $N_{X}=N_{X}(a)<\infty $, defined for any $a\in (0,1)$, and
satisfying that for all $n>N_{X}$, 
\begin{equation*}
\left\Vert X_{n}\right\Vert _{\infty }\leq a\log n.
\end{equation*}%
The time $N_{X}$ is finite with probability one because $\left\Vert
X_{n}\right\Vert _{\infty }$ is well known to grow at rate $O_{p}\left( \log
\left( n\right) ^{1/2}\right) $ as $n\rightarrow \infty $.

The second is $N_{A}=N_{A}(\gamma )<\infty $ chosen for any given $\gamma
<E\left( A_{1}\right) $, satisfying that for $n>N_{A}$ 
\begin{equation}
A_{n}\geq \gamma n.  \label{eq:boundAgamma}
\end{equation}%
The time $N_{A}$ is finite with probability one because of the Strong Law of
Large Numbers.

The third is $N_{a}$ such that, for all $n>N_{a}$, we have 
\begin{equation}
n\gamma \geq A_{1}\,n^{a}\exp (\left\Vert X_{1}\right\Vert _{\infty }).
\label{eq:defNmu}
\end{equation}%
It is immediate that $N_{a}$ is finite almost surely because $a\in \left(
0,1\right) $.\noindent

By successively applying the preceding three displays, we find that for $%
n>N:=\max (N_{A},N_{X},N_{a})$ and any $t=t_{1},\ldots ,t_{d}$, we have%
\begin{eqnarray*}
-\log A_{n}+X_{n}(t) &\leq &-\log A_{n}+\left\Vert X_{n}\right\Vert _{\infty
} \\
&\leq &-\log A_{n}+a\log n \\
&\leq &-\log (n\gamma )+a\log n \\
&\leq &-\log A_{1}-\left\Vert X_{1}\right\Vert _{\infty }\leq -\log
A_{1}+X_{1}(t).
\end{eqnarray*}%
Therefore, we conclude that, for $t=t_{1},\ldots ,t_{d}$, 
\begin{equation*}
\sup_{n\geq 1}\left\{ -\log A_{n}+X_{n}(t)\right\} =\max_{1\leq n\leq
N}\left\{ -\log A_{n}+X_{n}(t)\right\} .
\end{equation*}

The work in \cite{LBDM2016} explains how to simulate the random variables $%
N_{X}$, $N_{A}$, and $N_{a}$, jointly with the sequence $(A_{n})_{n\leq N}$
as well as $(X_{n})_{n\leq N}$. Moreover, it is also shown in \cite{LBDM2016}
that the number of random variables required to simulate $N_{X}$, $N_{A}$
and $N_{a}$ (jointly with $X_{1},...,X_{N}$ and $A_{1},...,A_{N}$) has
finite moments of any order. Therefore, $N$ has finite moments of any order.
Moreover, $E(N)=O(d^{\epsilon })$ for any $\epsilon >0$. In the appendix, we
reproduce the simulation procedure developed in \cite{LBDM2016}.

Now, observe that conditional on $X_{1},....,X_{N_{X}},N_{X}$, for $n>N_{X}$
the random vectors $\left( X_{k}\right) _{k\geq n}$ are independent, but
they no longer the follow a Gaussian distribution. Nevertheless, the $X_{k}$%
s still have zero conditional means given that $n>N$. This is because 
\begin{eqnarray*}
&&E\left( X_{n}\text{ }|\text{\ }\left\Vert X_{n}\right\Vert _{\infty }\leq
a\log n\right)  \\
&=&E\left( -X_{n}\text{ }|\text{\ }\left\Vert -X_{n}\right\Vert _{\infty
}\leq a\log n\right) =E\left( -X_{n}\text{ }|\text{\ }\left\Vert
X_{n}\right\Vert _{\infty }\leq a\log n\right) .
\end{eqnarray*}%
Consequently, we have that%
\begin{equation*}
E\left( e_{i}^{T}\Sigma ^{-1}X_{n}|n>N\right) =0.
\end{equation*}%
Therefore, because $M$ is independent of $X_{n}$ conditional on $n>N$, we
obtain that 
\begin{eqnarray}
&&E\left( G_{i}\left( x-M\right) \cdot e_{i}^{T}\Sigma ^{-1}X_{n}|n>N\right) 
\label{Cond_Ind} \\
&=&E\left( G_{i}\left( x-M\right) |n>N\right) \cdot E\left( e_{i}^{T}\Sigma
^{-1}X_{n}|n>N\right) =0.  \notag
\end{eqnarray}

One can let $n\rightarrow \infty $ in (\ref{Est_fn}) and formally apply (\ref%
{Cond_Ind}) leading to the following result, which is rigorously established
in \cite{BL2017}.

\begin{proposition}
For any $\left( x_{1},...,x_{d}\right) \in R^{d}$,%
\begin{equation}
f\left( x_{1},...,x_{d}\right) =-E\left(
\sum_{i=1}^{d}\sum_{k=1}^{N}G_{i}\left( x-M\right) \cdot e_{i}^{T}\Sigma
^{-1}X_{k}\right) .  \label{Eq_fAX}
\end{equation}
\end{proposition}

\subsection{Step 4: Variance Control in Malliavin-Thalmaier Estimators\label%
{Sec_Mall_Var}}

We now explain how to address the second issue discussed in Section \ref%
{Sec_Malliavin_Intro}, namely, controlling the variance when using the
Malliavin-Thalmaier estimator (\ref{Eq_fAX}).

Let us write%
\begin{equation*}
W\left( x\right) =-\sum_{i=1}^{d}\sum_{k=1}^{N}G_{i}\left( x-M\right) \cdot
e_{i}^{T}\Sigma ^{-1}X_{k},
\end{equation*}%
and observe that%
\begin{equation*}
W\left( x\right) =\frac{\left\langle M-x,\sum_{i=1}^{N}\Sigma
^{-1}X_{k}\right\rangle }{dw_{d}||M-x||^{d}}.
\end{equation*}

It turns out that the variance of $W\left( x\right) $ blows up because of
the singularity in the denominator when $M=x$. This is verified in \cite%
{BL2017}, but a similar calculation is also given in the setting of
diffusions in \cite{KY2009}. So, instead we consider an approximating
sequence defined via $\bar{W}_{0}\left( x\right) =0$, and%
\begin{equation*}
\bar{W}_{n}\left( x\right) =\frac{\left\langle M-x,\sum_{i=1}^{N}\Sigma
^{-1}X_{k}\right\rangle }{dw_{d}||M-x||^{d}+dw_{d}\delta _{n}||M-x||},\quad
n\geq 1,
\end{equation*}%
where 
\begin{equation*}
\delta _{n}=1/\log \log \log \left( n+e^{e}\right) .
\end{equation*}

It is immediate that $\lim_{n\rightarrow \infty }\bar{W}_{n}\left( x\right)
=W\left( x\right) $ almost surely. The use of a perturbation in the
denominator of the Malliavin-Thalmaier estimator is not new. In \cite{KY2009}
also a small positive perturbation in the denominator is added, but such
perturbation is, in their case, deterministic. The difference here is that
our perturbation contains the factor $\delta_n \left\Vert M-x \right\Vert$.
We have chosen our perturbation in order to ultimately control both the
variance and the bias of our estimator.

In order to quickly motivate the variance implications of our choice note
that 
\begin{equation*}
\left\vert \frac{\left\langle M-x,\sum_{i=1}^{N}\Sigma
^{-1}X_{k}\right\rangle }{dw_{d}||M-x||^{d}+dw_{d}\delta _{n}||M-x||}%
\right\vert \leq \left\vert \frac{\left\langle M-x,\sum_{i=1}^{N}\Sigma
^{-1}X_{k}\right\rangle }{dw_{d}\delta _{n}||M-x||}\right\vert \leq \frac{1}{%
dw_{d}\delta _{n}}\left\Vert \sum_{i=1}^{N}\Sigma ^{-1}X_{k}\right\Vert _{2},
\end{equation*}%
leading to a bound that does not explicitly contain $M$. Moreover, we
mentioned before that $N$ has finite moments of any order and $X_{k}$ is
Normally distributed, therefore, one can easily verify that $\left\Vert
\sum_{i=1}^{N}\Sigma ^{-1}X_{k}\right\Vert _{2}$ has finite moments of any
order, in particular finite second moment and therefore $\bar{W}_{n}\left(
x\right) $ has finite variance.

The reader might wonder why choosing $\delta _{n}$ in the definition of $%
\bar{W}_{n}\left( x\right) $, since any function of $n$ decreasing to zero
will ensure the convergence almost surely of $\bar{W}_{n}\left( x\right) $
towards $W\left( x\right) $. The previous upper bound, although not sharp
when $n$ is large, might also hint to the fact that is desirable to choose a
slowly varying function of $n$ in the denominator (at least the reader
notices a bound which deteriorates slowly as $n$ grows).

The precise reason for the selection of our perturbation in the denominator
obeys to a detailed variance calculation which can be seen in \cite{BL2017}.
A more in-depth discussion is given in Section \ref{Section_Estimator}
below. For the moment, let us continue with our development in order to give
the final form of our estimator.

Even though $\bar{W}_{n}\left( x\right) $ has finite variance and is close
to $W\left( x\right) $, unfortunately, we have that $\bar{W}_{n}\left(
x\right) $ is no longer an unbiased estimator of $f\left( x\right) $. In
order to remove the bias we take advantage of a randomization idea from \cite%
{RG2015} and \cite{McL2011}, which is related to the multilevel Monte Carlo
method in \cite{Giles2008}, as we shall explain next.

\subsection{Final Form of Our Estimator\label{Section_Estimator}}

Let us define $\bar{W}_{0}\left( x\right) =0$ and for $n\geq 1$ let us write%
\begin{equation*}
\Delta _{n}\left( x\right) =\bar{W}_{n}\left( x\right) -\bar{W}_{n-1}\left(
x\right) .
\end{equation*}%
In order to facilitate the variance analysis of our randomized multilevel
Monte Carlo estimator we further consider a sequence $\left( \bar{\Delta}%
_{n}\left( x\right) \right) _{n\geq 1}$ of independent random variables so
that $\Delta _{n}\left( x\right) $ and $\bar{\Delta}_{n}\left( x\right) $
are equal in distribution.

We let $L$ be a random variable taking values on $n\geq 1$, independent of
everything else. Moreover, we let $g\left( n\right) =P\left( L\geq n\right) $
and assume that 
\begin{equation*}
g\left( n\right) = n^{-1}\left( \log \left( n+e-1\right) \right) ^{-1}\left(
\log\left( \log \left( n+e^e-1\right) \right)\right) ^{-1}.
\end{equation*}

Then, the final form of our estimator is 
\begin{equation}
V\left( x\right) =\sum_{k=1}^{L}\frac{\bar{\Delta}_{k}\left( x\right) }{%
g\left( k\right) }.  \label{Final_Estimator}
\end{equation}%
The estimator $V\left( x\right) $ can be easily simulated assuming that we
can sample $M$ exactly, jointly with $X_{1},...,X_{N},N$. This will be
explained in \textbf{Algorithm M} in Section \ref{Sec_Sample_M_all}.

The choice of $g\left( \cdot \right) $ and the selection of the factor $%
\delta _{n}$ appearing in the denominator of $\bar{W}_{n}\left( x\right) $
are closely related. In the end, the randomized multilevel Monte Carlo idea
applied formally yields that 
\begin{eqnarray}  \label{eqn_V_excg}
E\left( V\left( x\right) \right) &=&E\left( \sum_{k=1}^{\infty }\frac{\bar{%
\Delta}_{k}\left( x\right) I\left( L\geq k\right) }{g\left( k\right) }%
\right) =\sum_{k=1}^{\infty }E\left( \frac{\bar{\Delta}_{k}\left( x\right)
I\left( L\geq k\right) }{g\left( k\right) }\right)  \label{Bias_0} \\
&=&\sum_{k=1}^{\infty }E\left( \bar{\Delta}_{k}\left( x\right) \right)
=E\left( W\left( x\right) \right) -E\left( \bar{W}_{0}\left( x\right)
\right) =E\left( W\left( x\right) \right) =f\left( x\right) .  \notag
\end{eqnarray}

In order to make the previous manipulations rigorous, we must justify
exchanging the summation in (\ref{eqn_V_excg}). In turn, it suffices to make
sure that $\sum_{k\geq 1}E\left( \left\vert \bar{\Delta}_{k}\left( x\right)
\right\vert \right) <\infty $. In addition, we also need to guarantee that $%
V\left( x\right) $ has finite variance. These and other properties will be
used to obtain confidence intervals for our estimates given a computational
budget. We shall summarize the properties of $V(x)$ in our main result given
in the next section, which also provides a discussion of the running time
analysis which motivates the choice of $g\left( n\right) $.

\section{Main Result\label{Sect_Main}}

Our main contribution is summarized in the following result, which is fully
proved in \cite{BL2017}. Our objective now is to sketch the gist of the
technical development in order to have at least an intuitive understanding
of the choices behind the design of our estimator (\ref{Final_Estimator}).
We measure computational cost in terms of the elementary random variables
simulated.

\begin{theorem}
\label{Thm_Main}Let $\varrho $ be the cost required to regenerate $M$ so
that $V\left( x\right) $, defined in (\ref{Final_Estimator}), has a
computational cost equal to $C=\sum_{i=1}^{L}\varrho _{i}+1$ (where $L$ is
independent of $\varrho _{1},\varrho _{2},\ldots $, which are i.i.d. copies
of $\varrho $). Let $\left( V_{1}\left( x\right) ,C_{1}\right) ,(V_{2}\left(
x\right) ,C_{2}),...$ be i.i.d. copies of $\left( V\left( x\right) ,C\right) 
$ and set $T_{n}=C_{1}+...+C_{n}$ with $T_{0}=0$. For each $b>0$ define, $%
B\left( b\right) =\max \{n\geq 0:T_{n}\leq b\}$, then we have that%
\begin{equation}
f\left( x\right) =E\left( V\left( x\right) \right) \text{ \ and }Var\left(
V\left( x\right) \right) <\infty .  \label{Mean_Variance}
\end{equation}%
Moreover, 
\begin{equation*}
\sqrt{\frac{b}{E\left( \varrho _{1}\right) \cdot \log \log \log \left(
b\right) }}\left( \frac{1}{B\left( b\right) }\sum_{i=1}^{B\left( b\right)
}V_{i}\left( x\right) -f\left( x\right) \right) \Rightarrow N\left(
0,Var\left( V\left( x\right) \right) \right) .
\end{equation*}
\end{theorem}

Before we discuss the analysis of the proof of Theorem \ref{Thm_Main}, it is
instructive to note that the previous result can be used to obtain
confidence intervals for the value of the density $f\left( x\right) $ with
precision $\varepsilon $ at a computational cost of order $O\left(
\varepsilon ^{-2}\log \log \log \left( 1/\varepsilon \right) \right) $,
given a fix confidence level (see Section \ref{Section_Numerics} for an
example of how to produce such confidence interval).

The quantity $B\left( b\right) $ denotes the number of i.i.d. copies of $%
V\left( x\right) $ which can be simulated with a computational budget $b$,
so the pointwise estimator given in Theorem \ref{Thm_Main} simply is the
empirical average of $B\left( b\right) $ i.i.d. copies of $V\left( x\right) $%
. 

The rate of convergence implied by Theorem \ref{Thm_Main} is, for all
practical purpose, the same as the highly desirable canonical rate $O\left(
\varepsilon ^{-2}\right) $, which is rarely achieved in complex density
estimation problems, such as the one that we consider in this paper.

\subsection{Sketching the Proof of Theorem \protect\ref{Thm_Main}\label%
{Subsec_Pf_Thm_1}}

At the heart of the proof of Theorem \ref{Thm_Main} lies a bound on the size
of $\left\vert \Delta _{n}\left( x\right) \right\vert $. For notational
simplicity, let us concentrate on $\left\vert \Delta _{n}\left( 0\right)
\right\vert $ and note that for any $\beta \geq 1$%
\begin{eqnarray}
\left\vert \Delta _{n}\left( 0\right) \right\vert ^{\beta } &\leq &\frac{%
\left\Vert \Sigma ^{-1}\right\Vert ^{\beta }}{\left( dw_{d}\right) ^{\beta }}%
\left( \sum_{i=1}^{N}\left\Vert X_{k}\right\Vert \right) ^{\beta }
\label{Del_n} \\
&&\times \left\vert \frac{\left\Vert M\right\Vert }{||M||^{d}+||M||\delta
_{n+1}}-\frac{\left\Vert M\right\Vert }{||M||^{d}+||M||\delta _{n}}%
\right\vert ^{\beta }.  \notag
\end{eqnarray}%
We have argued that, because $N$ has finite moments of any order, the random
variable $\sum_{i=1}^{N}\left\Vert X_{k}\right\Vert _{2}$ is easily seen to
have finite moments of any order. So, after applying H\"{o}lder's inequality
to the right hand side of (\ref{Del_n}), it suffices to concentrate on
estimating, for any $q>1$,%
\begin{eqnarray*}
&&E\left( \left\vert \frac{1}{||M||^{d-1}+\delta _{n+1}}-\frac{1}{%
||M||^{d-1}+\delta _{n}}\right\vert ^{\beta q}\right) ^{1/q} \\
&=&E\left( \left\vert \frac{\delta _{n}-\delta _{n+1}}{\left(
||M||^{d-1}+\delta _{n+1}\right) \left( ||M||^{d-1}+\delta _{n}\right) }%
\right\vert ^{\beta q}\right) ^{1/q}.
\end{eqnarray*}%
Let us define 
\begin{equation}
a\left( n\right) :=\delta _{n}-\delta _{n+1}\sim \delta _{n}^{2}\frac{1}{%
\log \log \left( n\right) \cdot \log \left( n\right) \cdot n},
\label{Def_a_n}
\end{equation}%
and focus on 
\begin{equation}
D_{n,\beta }\left( 0\right) :=E\left( \left\vert \frac{1}{\left(
||M||^{d-1}+\delta _{n+1}\right) \left( ||M||^{d-1}+\delta _{n}\right) }%
\right\vert ^{\beta q}\right) ^{1/q}.  \label{D_n0}
\end{equation}

Assuming that $M$ has a continuous density in a neighborhood of the origin
(a fact which can be shown, for example, from (\ref{ID_JOIN}), using the
Gaussian property of the $X_{n}$s), we can directly analyze (\ref{D_n0})
using a polar coordinates transformation, obtaining that for some $\kappa >0$%
\begin{equation}
D_{n,\beta }^{q}\left( 0\right) \leq \kappa \int_{0}^{\infty }\int_{\theta
\in \mathcal{S}^{d-1}}\frac{f\left( r\cdot \theta \right) r^{d-1}}{\left(
r^{d-1}+\delta _{n+1}\right) ^{\beta q}\left( r^{d-1}+\delta _{n}\right)
^{\beta q}}drd\theta ,  \label{Dn0_0}
\end{equation}%
where $\mathcal{S}^{d-1}$ represents the surface of the unit ball in $d$
dimensions. Further study of the decay properties of $f\left( r\cdot \theta
\right) $ as $r$ grows large, uniformly over $\theta \in \mathcal{S}^{d-1}$,
allows us to conclude that 
\begin{equation}
D_{n,\beta }^{q}\left( 0\right) \leq \kappa ^{\prime }\int_{0}^{\infty }%
\frac{r^{d-1}}{\left( r^{d-1}+\delta _{n+1}\right) ^{\beta q}\left(
r^{d-1}+\delta _{n}\right) ^{\beta q}}dr,  \label{Dn_0_1}
\end{equation}%
for some $\kappa ^{\prime }>0$. Applying the change of variables $r=u\delta
_{n}^{1/(d-1)}$ to the right-hand side of (\ref{Dn_0_1}), allows us to
conclude, after elementary algebraic manipulations that%
\begin{equation*}
D_{n,\beta }^{q}\left( 0\right) =O\left( \delta _{n}^{d/(d-1)-2\beta
q}\right) ,
\end{equation*}%
therefore concluding that 
\begin{equation}
E\left( \left\vert \Delta _{n}\left( 0\right) \right\vert ^{\beta }\right)
=O\left( \left( \frac{\delta _{n}-\delta _{n+1}}{\delta _{n}^{2}}\right)
^{\beta }\delta _{n}^{d/(q(d-1))-2\beta }\right) .  \label{Del_n_beta}
\end{equation}%
Setting $\beta =1$ we have (from (\ref{Def_a_n}) and the definition of $%
\delta _{n}$) that 
\begin{equation}
\sum_{n\geq 1}E\left( \left\vert \Delta _{n}\left( 0\right) \right\vert
\right) =O\left( \sum_{n\geq 1}\frac{1}{\log \log \left( n\right) \cdot \log
\left( n\right) \cdot n}\delta _{n}^{d/(q(d-1))}\right) <\infty ,
\label{Bnd_EV}
\end{equation}%
because $d/(d-1)>1$ and $q>1$ can be chosen arbitrarily close to one. This
estimate justifies the formal development in (\ref{Bias_0}) and the fact
that $EV\left( x\right) =f\left( x\right) $.

Now, the analysis in \cite{RG2015} states that $Var\left( V\left( x\right)
\right) <\infty $ if 
\begin{equation}
\sum_{n\geq 1}\frac{E\left\vert \bar{\Delta}_{n}\left( 0\right) \right\vert
^{2}}{g\left( n\right) }<\infty .  \label{Bnd_Var_RG}
\end{equation}%
Once again, using (\ref{Del_n_beta}) and our choice of $g\left( n\right) $,
we obtain that (\ref{Bnd_Var_RG}) holds because of the estimate 
\begin{equation}
\sum_{n\geq 1}\frac{n\cdot \log \left( n\right) \cdot \log \log \left(
n\right) }{\left( \log \log \left( n\right) \cdot \log \left( n\right) \cdot
n\right) ^{2}}\delta _{n}^{d/(q(d-1))}<\infty ,  \label{D_n2_div_G}
\end{equation}%
which is, after immediate cancellations, completely analogous to (\ref%
{Bnd_EV}).

Finally, because the cost of sampling $M$ (in terms of the number of
elementary random variables, such as multivariate Gaussian random variables)
has been shown to have finite moments of any order \cite{LBDM2016}, one can
use standard results from the theory of regular variation (see \cite{RS2008}%
) to conclude that 
\begin{equation*}
P\left( \sum_{i=1}^{L}\varrho _{i}+1>t\right) \sim P\left( L>t/E\left(
\varrho _{1}\right) \right) \sim E\left( \varrho _{1}\right) t^{-1}\log
\left( t\right) ^{-1}\log \log \left( t\right) ^{-1},
\end{equation*}

as $t\rightarrow \infty $. Now, the form of the Central\ Limit Theorem is an
immediate application of Theorem 1 in \cite{ZBG2017}.

\section{Numerical Examples\label{Section_Numerics}}

In this section, we implement our estimator and compare it against a
conventional kernel density estimator. We measure the computational cost in
terms of the number of independent samples drawn from \textbf{Algorithm M}.
This convention translates into assuming that $E(\rho _{1})=1$ in Theorem~%
\ref{Thm_Main}. Given a computational budget $b$, the estimated density is
given by 
\begin{equation*}
\hat{f}_{b}(x)=\frac{\sum_{i=1}^{B(b)}V_{i}\left( x\right) }{B(b)}.
\end{equation*}

According to Theorem~\ref{Thm_Main}, we can construct the confidence
interval for underlying density $f(x)$ with significance level $\alpha $ as 
\begin{equation*}
\left( \hat{f}_{b}(x)-z_{\alpha /2}\hat{s}\sqrt{a(b)},\hat{f}%
_{b}(y)+z_{\alpha /2}\hat{s}\sqrt{a(b)}\right) ,
\end{equation*}%
where $z_{\alpha /2}$ is the quantile corresponding to the $1-\alpha /2$
percentile, 
\begin{equation*}
\hat{s}^{2}=\frac{\sum_{i=1}^{B(b)}\left( V_{i}\left( x\right) -\hat{f}%
_{b}(x)\right) ^{2}}{B(b)},
\end{equation*}
and%
\begin{equation*}
a(b)=\sqrt{\frac{\log \log \log \left( b\right) }{b}}.
\end{equation*}

\begin{figure}[h]
\label{fig_denEst} \centering
\includegraphics[width = 0.5\textwidth]{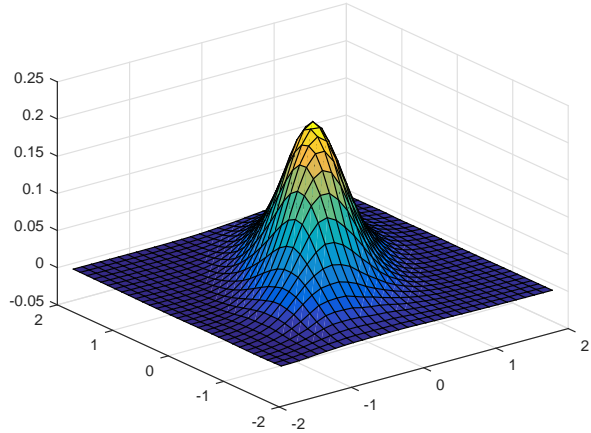}
\caption{The estimated 3-dimensional joint density of a max-stable process
using our algorithm}
\end{figure}

We perform our algorithm to estimate the density of the max-stable process.
We assume that $T=[0,1]$ and $X_{n}(\cdot )$ is a standard Brownian motion.
We are interested in estimating the density of $M=\left(
M(1/3),M(2/3),M(1)\right) ^{T}$. That is, the spatial grid is $(1/3,2/3,1)$.
The graph in Figure~\ref{fig_denEst} shows a plot of the density on the set $%
\{x\in \mathbb{R}^{3}:x_{1}\in (-2,2),x_{2}\in (-2,2),x_{3}=0\}$. Our
estimation of this 3-dimensional density has a computation budget of $%
B=10^{6}$ samples from \textbf{Algorithm M}. 

We calculate the $95\%$ confidence interval of the density on several
selected values of the process $M(\cdot)$.

\begin{center}
\begin{tabular}{|l|l|l|l|l|l|}
\hline
Values ($x$) & (0,0,0) & (0,0.5,0) & (0.5,0,0) & (0,-0.5,0) & (-0.5,0,0) \\ 
\hline
est. density ${\hat{f}}_{b}{(x)}$ & 0.2126 & 0.106 & 0.1292 & 0.1039 & 0.1439
\\ \hline
lower CI & 0.1916 & 0.0971 & 0.1180 & 0.0947 & 0.1311 \\ \hline
upper CI & 0.2336 & 0.1149 & 0.14036 & 0.1131 & 0.1567 \\ \hline
Relative error & 5.05\% & 4.29\% & 4.41\% & 4.54\% & 4.53\% \\ \hline
\end{tabular}
\end{center}

As a comparison, we also calculate the $95\%$ confidence interval of the
density using the plug-in kernel density estimation (KDE) method with the
same amount ($b=10^{6}$) of i.i.d. samples of $M$. We use the normal density
function as the kernel function and select the bandwidth according to \cite%
{SS2005}. The estimator is obtained as follows. Sample $M^{(1)},M^{(2)},%
\ldots ,M^{(d)}$ i.i.d. copies of $M$, let $h_{b}=b^{-1/(2d+1)}$ and compute
the sample covariance matrix, $\hat{\Sigma}$, based on $(M^{(1)},M^{(2)},%
\ldots ,M^{(d)})$. Then, let 
\begin{equation*}
\hat{f}_{b}^{KDE}(x)=\frac{1}{bh_{b}^{d}}\sum_{i=1}^{b}\phi \left( A^{-\frac{%
1}{2}}\frac{x-M^{\left( i\right) }}{h_{b}}\right) ,
\end{equation*}%
where $A=\hat{\Sigma}/$det$|\hat{\Sigma}|$. We apply the method from \cite%
{Fiorio2004} to evaluate the corresponding confidence interval, thereby
obtaining the following estimates,

\begin{center}
\begin{tabular}{|l|l|l|l|l|l|}
\hline
Values ($x$) & (0,0,0) & (0,0.5,0) & (0.5,0,0) & (0,-0.5,0) & (-0.5,0,0) \\ 
\hline
est. density $\hat{f}_{b}^{KDE}(x)$ & 0.2163 & 0.0846 & 0.1143 & 0.0938 & 
0.1084 \\ \hline
lower CI & 0.1953 & 0.0712 & 0.0999 & 0.0800 & 0.0934 \\ \hline
upper CI & 0.2373 & 0.0980 & 0.1287 & 0.1076 & 0.1234 \\ \hline
Relative error & 4.94\% & 8.07\% & 6.43\% & 7.51\% & 7.05\% \\ \hline
\end{tabular}
\end{center}

From the above tables, we can see that our algorithm provides similar
estimates to those obtained using the KDE. However, our estimator also has a
smaller relative error when the estimated value is relatively small. Also,
as discussed in \cite{Fiorio2004}, one must carefully choose the bandwidth
to guarantee coverage because the KDE may be asymptotically biased. In
contrast, the construction of confidence intervals with our estimator is a
straightforward application of elementary statistical tools. 
%Moreover, we can have an
%unbiased estimation if we use fixed number of estimator samples instead of
%fixed computational budgets, while the asymptotic bias in KDE is difficult
%to get. Another advantage of our method is that the explicit form of our
%estimator could be used in the statistical inference of the underlying
%process, given the available data of the max-stable process. In conclusion,
%from the above arguments, we developed an estimator superior to the KDE in
%terms of the application in max-stable processes.

\section{Appendix: A Detailed Algorithmic Implementation\label{Sec_App}}

In order to make this paper as self-contained as possible, we reproduce here
the algorithms from \cite{LBDM2016} which allow us to simulate the random
variables $N_{X}$, $N_{A}$, and $N_{a}$, jointly with $(A_{n})_{n\leq N}$
and $(X_{n})_{n\leq N}$.

\subsection{Simulating Last Passage Times of Random Walks}

Define the random walk $S_{n}=\gamma n-A_{n}$ for $n\geq 0$. Note that $%
ES_{n}<0$, by our choice of $\gamma <E\left( A_{1}\right) $. The authors in 
\cite{LBDM2016}, argue that the choice of $\gamma $ is not too consequential
so we shall assume that $\gamma =1/2$.

Here we review an algorithm from \cite{LBDM2016} for finding a random time $%
N_{S}$ such that $S_{n}<0$ for all $n>N_{S}$. Observe that $N_{S}=N_{A}$.

The algorithm is based on alternately sampling upcrossings and downcrossings
of the level $0$. We write $\xi _{0}^{+}=0$ and, for $i\geq 1$, we
recursively define 
\begin{equation*}
\xi _{i}^{-}=%
\begin{cases}
\inf \{n\geq \xi _{i-1}^{+}:S_{n}<0\} & \text{if }\xi _{i-1}^{+}<\infty  \\ 
\infty  & \text{otherwise}%
\end{cases}%
\end{equation*}%
together with 
\begin{equation*}
\xi _{i}^{+}=%
\begin{cases}
\inf \{n\geq \xi _{i}^{-}:S_{n}\geq 0\} & \text{if }\xi _{i}^{-}<\infty  \\ 
\infty  & \text{otherwise}.%
\end{cases}%
\end{equation*}%
As usual, in these definitions, the infimum of an empty set should be
interpreted as $\infty $. Writing 
\begin{equation*}
N_{S}=\sup \{\xi _{n}^{-}:\xi _{n}^{-}<\infty \},
\end{equation*}%
we have by construction $S_{n}<0$ for $n>N_{S}$. The random variable $N_{S}-1
$ is an upward last passage time: 
\begin{equation*}
N_{S}-1=\sup \{n\geq 0:S_{n}\geq 0\}.
\end{equation*}%
Note that $0\leq N_{S}<\infty $ almost surely under $P$ since $%
(S_{n})_{n\geq 0}$ starts at the origin and has negative drift. We will
provide pseudo-codes for simulating $(S_{1},\ldots ,S_{N_{S}+\ell })$ for
any fixed $\ell \geq 0$, but first we need a few definitions.

First, we assume that the \emph{Cram\'{e}r's root}, $\theta >0$, satisfying $%
E(\exp (\theta S_{1}))=1$ has been computed. We shall use $\mathbb{P}_{x}$
to denote the measure under which $\left( A_{n}\right) _{n\geq 1}$ are
arrivals of a Poisson process with unit rate and $S_{0}=x$. Then, we define $%
P_{x}^{\theta }$ through an exponential change of measure. In particular, on
the $\sigma $-field generated by $S_{1},\ldots ,S_{n}$ we have 
\begin{equation*}
\frac{dP_{x}}{dP_{x}^{\theta }}=\exp (-\theta (S_{n}-x)).
\end{equation*}%
It turns out that under $P_{x}^{\theta }$, $\left( A_{n}\right) _{n\geq 1}$
corresponds to the arrivals of a Poisson process with rate $1-\theta $ and
the random walk $(S_{n})_{n\geq 1}$ has a positive drift.

To introduce the algorithm to sample $(S_{1},\ldots ,S_{N_{S}+\ell })$ we
first need the following definitions: 
\begin{equation*}
\tau ^{-}=\inf \{n\geq 0:S_{n}<0\},\quad \quad \tau ^{+}=\inf \{n\geq
0:S_{n}\geq 0\}.
\end{equation*}

For $x\geq 0$, it is immediate that we can sample a downcrossing segment $%
S_{1},\ldots ,S_{\tau ^{-}}$ under $P_{x}$ due to the negative drift, and we
record this for later use in a pseudocode function. \emph{Throughout our
discussion,`sample' in pseudocode stands for `sample independently of
anything that has been sampled already'.}

\textbf{Function }\textsc{SampleDowncrossing}\textbf{(}$x$\textbf{): Samples 
}$(S_{1},\ldots ,S_{\tau ^{-}})$\textbf{\ under }$P_{x}$\textbf{\ for }$%
x\geq 0$

Step 1:\qquad Return sample $S_{1},\ldots ,S_{\tau ^{-}}$ under $P_{x}$.

Step 2: EndFunction

\bigskip

Sampling an upcrossing segment is more interesting because it is possible
that $\tau ^{+}=\infty $. So, an algorithm needs to be able to detect this
event within a finite amount of computing resources. For this reason, we
understand sampling an upcrossing segment under $P_{x}$ for $x<0$ to mean
that an algorithm outputs $S_{1},\ldots ,S_{\tau ^{+}}$ if $\tau ^{+}<\infty 
$, and otherwise it outputs `degenerate'. The following pseudo-code samples
an upcrossing under $P_{x}$ for $x<0$.

\bigskip

\textbf{Function} \textsc{SampleUpcrossing}($x$): \textbf{Samples }$%
(S_{1},\ldots ,S_{\tau ^{+}})$\textbf{\ under }$P_{x}$\textbf{\ for }$x<0$

Step 1: $S\leftarrow $ sample $S_{1},\ldots ,S_{\tau ^{+}}$ under $%
P_{x}^{\theta }$

Step 2: $U\leftarrow $ sample a standard uniform random variable

Step 3: If {$U\leq \exp (-\theta (S_{\tau ^{+}}-x))$}

Step 4:\qquad Return $S$

Step 5: Else

Step 6:\qquad Return `degenerate'

Step 7: EndIf

Step 8: EndFunction

%It is worth noting that, as a special case, this is a sampling algorithm for $\1(\tau^+<\infty)$ under $\mathbb{P}_x$.

\bigskip

We next describe how to sample $(S_{k})_{k=1,\ldots ,n}$ from $P_{x} $
conditionally on $\tau ^{+}=\infty $ for $x<0$. Since $\tau ^{+}=\infty $ is
equivalent to $\sup_{k\leq \ell }S_{k}<0$ and $\sup_{k>\ell }S_{k}<0$ for
any $\ell \geq 1$, after sampling $S_{1},\ldots ,S_{\ell }$, by the Markov
property we can use \textsc{SampleUpcrossing}$(S_{\ell })$ to verify whether
or not $\sup_{k>\ell }S_{k}<0$.

\bigskip

\textbf{Function }\textsc{SampleWithoutRecordS}$\left( {x,\ell }\right) $%
\textbf{: Samples }$(S_{k})_{k=1,\ldots ,\ell }$\textbf{\ from }$P_{x}$%
\textbf{\ given }$\tau ^{+}=\infty $\textbf{\ for }$\ell \geq 1$\textbf{, }$%
x<0$

Step 1: Repeat

Step 2:$\qquad S\leftarrow $ sample $(S_{k})_{k=1,\ldots ,\ell }$ under $%
P_{x}$

Step 3: Until $\sup_{1\leq k\leq \ell }S_{k}<0$ and \textsc{SampleUpcrossing}%
$(S_{\ell })$ is `degenerate'

Step 4: Return $S$

Step 5: EndFunction

\bigskip

We summarize our discussion with the full algorithm for sampling $%
(S_{0},\ldots ,S_{N_{S}+\ell })$ under $P$ given some $\ell \geq 0$.

\subparagraph{\textbf{Algorithm S: Samples }$S=(S_{0},\ldots ,S_{N_{S}+\ell
})$\textbf{\ under }$P$\textbf{\ for }$\ell \geq 0$}

\label{alg:N_A}

\# We use $S_{\text{end}}$ to denote the last element of $S$.

Step 1: $S\leftarrow \lbrack 0]$

Step 2: \ Repeat

Step 3:\qquad DowncrossingSegment $\leftarrow $ \textsc{SampleDowncrossing}$%
(S_{\text{end}})$

Step 4:$\qquad S\leftarrow \lbrack S,$Do$\text{wncrossingSegment}]$

Step 5:\qquad UpcrossingSegment $\leftarrow $\textsc{SampleUpcrossing}$(S_{%
\text{end}})$

Step 6:\qquad If{\ UpcrossingSegment is not `degenerate'}

Step 7:$\qquad \qquad S\leftarrow \lbrack S,\text{upcrossingSegment}]$

Step 8: \qquad EndIf

Step 9: \ \ Until {UpcrossingSegment is `degenerate'}

Step 10: If {$\ell >0$}

Step 11: $\qquad S\leftarrow \lbrack S,$\textsc{SampleWithoutRecordS}$(S_{%
\text{end}},\ell )]$

Step 12: EndIf

\subsection{Simulating Last Passage Times for Maxima of Gaussian Vectors}

The technique is similar to the random walk case using a sequence of
record-breaking times. The parameter $a\in \left( 0,1\right) $ can be chosen
arbitrarily, but \cite{LBDM2016} suggests selecting $a$ such that 
\begin{equation*}
\exp \left( \frac{\overline{\sigma }}{a}\overline{\Phi }^{-1}\left( \delta 
\sqrt{2\pi }{\frac{\phi (\overline{\sigma }/a)}{d\overline{\sigma }/a}}%
\right) +\frac{\overline{\sigma }^{2}}{a^{2}}\right) =E\left[ \left( \frac{%
A_{1}\,\exp (\left\Vert X\right\Vert _{\infty })}{\gamma }\right) ^{\frac{1}{%
1-a}}\right] ,
\end{equation*}%
where $\Phi (\cdot )$ is the cumulative distribution function of a standard
Gaussian random variable and $\overline{\Phi }=1-\Phi $.

Now, assume that $\eta _{0}\geq 0$ is given (we will choose it specifically
in the sequel). Let $(X_{n})_{n\geq 1}$ be i.i.d. copies of $X$ and define,
for $i\geq 1$, a sequence of \emph{record breaking times} $(\eta _{i})$
through 
\begin{equation*}
\eta _{i}=%
\begin{cases}
\inf \{n>\eta _{i-1}:\left\Vert X_{n}\right\Vert _{\infty }>a\log n\} & 
\text{if }\eta _{i-1}<\infty \\ 
\infty & \text{otherwise.}%
\end{cases}%
.
\end{equation*}

We provide pseudo-codes which ultimately will allow us to sample $%
(X_{1},\ldots ,X_{N_{X}+\ell })$ for any fixed $\ell \geq 0$, where 
\begin{equation*}
N_{X}=\max \{\eta _{i}:\eta _{i}<\infty \}.
\end{equation*}%
First, we shall discuss how to sample $(X_{n})$ up to a $\eta _{1}$. In
order to sample $\eta _{1}$, $\eta _{0}=n_{0}$ needs to be chosen so that $%
P(\left\Vert X\right\Vert _{\infty }>a\log n)$ is controlled for every $%
n>n_{0}$. Given the choice of $a\in \left( 0,1\right) $, select $n_0$ such
that 
\begin{equation*}
{d\overline{\Phi }\left( \frac{a\log n_{0}}{\overline{\sigma }}-\frac{%
\overline{\sigma }}{a}\right) }\leq \frac{1}{2}\sqrt{\frac{\pi }{2}}{\frac{%
\phi (\overline{\sigma }/a)}{\overline{\sigma }/a}}.
\end{equation*}

Define 
\begin{equation}
T_{n_{0}}=\inf \{k\geq 1:\left\Vert X_{k}\right\Vert _{\infty }>a\log
(n_{0}+k)\}.  \label{Def_T_n0}
\end{equation}%
We describe an algorithm that outputs `degenerate' if $T_{n_{0}}=\infty $
and $(X_{1},\ldots ,X_{T_{n_{0}}})$ if $T_{n_{0}}<\infty $.

First, we describe a simple algorithm to simulate from $X$ conditioned on $%
\left\Vert X\right\Vert _{\infty }>a\log n$. Our algorithm makes use of a
probability measure $P^{(n)}$ defined through 
\begin{equation*}
\frac{dP^{(n)}}{dP}(x)=\frac{\sum_{i=1}^{d}\mathbf{1}(\left\vert
x(t_{i})\right\vert >a\,\log n)}{\sum_{i=1}^{d}{P(}\left\vert {X(t_{i})}%
\right\vert {>a\,\log n)}}.
\end{equation*}%
It turns out that the measure $P^{(n)}$ approximates the conditional
distribution of $X$ given that $\left\Vert X\right\Vert _{\infty }>a\log n$
for $n$ large.

Now, define $w^{j}(t)=\mathrm{Cov}(X(t),X(t_{j}))/\mathrm{Var}\left( X\left(
t_{j}\right) \right) $ and note that $X\left( \cdot \right) -w^{\nu }\left(
\cdot \right) X(t_{\nu })$ is independent of $X(t_{\nu })$ given $\nu $.
This property is used in \cite{LBDM2016} to show that the following
algorithm outputs from $P^{(n)}$. We will let $U$ be a uniform random
variable in $\left( 0,1\right) $ and $J$ is independent of $U$ and such that 
$P\left( J=1\right) =1/2=P\left( J=-1\right) $.)

\bigskip

\textbf{Function} \textsc{ConditionedSampleX} $\left( a,n\right) $\textbf{:
Samples $X$ from} $P^{(n)}$

Step 1: $\nu \leftarrow $ sample with probability mass function 
\begin{equation*}
P(\nu =j)=\frac{P(\left\vert X(t_{j})\right\vert >a\log n)}{\sum_{i=1}^{d}{P(%
}\left\vert {X(t_{i})}\right\vert {>a\log n)}}
\end{equation*}

Step 2: $U\gets $ sample a standard uniform random variable

Step 3: $X(t_{\nu })\leftarrow \sigma (t_{\nu })\cdot J\cdot \Phi
^{-1}\left( U+(1-U)\Phi \left( a\left( \log n\right) /\sigma (t_{\nu
})\right) \right) $ \# Conditions on $\left\vert X(t_{\nu })\right\vert
>a\log n$

Step 4: $Y\leftarrow $ sample of $X$ under $P$

Step 5: Return $Y(t)- w^\nu(t) Y(t_\nu) + X(t_\nu)$

Step 6: EndFunction

\bigskip

We now explain how \textsc{ConditionedSampleX} is used to sample $T_{n_{0}}$%
. Define, for $k\geq 1$, 
\begin{equation*}
g_{n_{0}}(k)=\frac{\int_{k-1}^{k}\phi ((a\log (n_{0}+s))/\overline{\sigma }%
)ds}{\int_{0}^{\infty }\phi ((a\log (n_{0}+s))/\overline{\sigma })ds},
\end{equation*}%
where $\phi (x)=d \Phi(x)/dx$. Note that $g_{n_{0}}(\cdot )\geq 0$ defines
the probability mass function of some random variable $K$. It turns out that
if $U\sim U\left( 0,1\right) $ then we can sample 
\begin{equation*}
K=\left\lceil \exp \left\{ \frac{\overline{\sigma }^{2}}{a^{2}}+\frac{%
\overline{\sigma }}{a}\overline{\Phi }^{-1}\left( U\,\overline{\Phi }\left( 
\frac{a\log n_{0}}{\overline{\sigma }}-\frac{\overline{\sigma }}{a}\right)
\right) \right\} -n_{0}\right\rceil .
\end{equation*}

The next function samples $(X_{1},\ldots ,X_{T_{n_{1}}})$ for $n_{1}\geq
n_{0}$.

\bigskip

\textbf{Function} \textsc{SampleSingleRecord} $\left( a,n_{0},n_{1}\right) $%
\textbf{: Samples} $(X_{1},\ldots ,X_{T_{n_{1}}})$ \textbf{for} $a\in
(0,1),n_{1}\geq n_{0}\geq 0$

Step 1: Sample $K$

Step 2: $[X_{1},\ldots ,X_{K-1}]\leftarrow $ i.i.d. sample from $P$

Step 3: $X_{K}$ $\leftarrow \text{ \textsc{ConditionedSampleX}}(a,n_{1}+K)$

Step 4: $U\gets$ sample a standard uniform random variable

Step 5: If $\left\Vert {X_{k}}\right\Vert _{\infty }${$\leq a\log (n_{1}+k)$
for $k=1,\ldots ,K-1$ and $U\,{g_{n_{0}}(K)}\leq {dP/dP^{(n_{1}+K)}(X_{K})}$}

Step 6: \qquad Return $(X_1,\ldots,X_K)$

Step 7: Else

Step 8: \qquad Return `degenerate'

Step 9: EndIf

Step 10: EndFunction

\bigskip

We next describe how to sample $(X_{k})_{k=1,\ldots ,n}$ conditionally on $%
T_{n_{0}}=\infty $. This is a simple task because the $X_{n}$s are
independent.

\bigskip

\textbf{Function} \textsc{SampleWithoutRecordX} $\left(n_1,\ell\right)$ 
\textbf{: Samples} $(X_k)_{k=1,\ldots,\ell}$ \textbf{conditionally on} $%
T_{n_1}=\infty$ for $\ell\ge 1$

Step 1: Repeat

Step 2: \qquad $X \gets$ sample $(X_k)_{k=1,\ldots,\ell}$ under $P$

Step 3: Until $\sup_{1\leq k\leq \ell }[X_{k}-a\log (n_{1}+k)]<0$ 
%and $\textproc{SampleSingleRecord}(n_1+\ell)$ is `degenerate'

Step 4: Return $X$

Step 5: EndFunction

\bigskip

We now can explain how to sample $(X_{1},\ldots ,X_{N_{X}+\ell })$ under $P$
given some $\ell \geq 0$. The idea is to successively apply \textsc{%
SampleSingleRecord} to generate the sequence $\left( \eta _{i}:i\geq
1\right) $ defined at the beginning of this section. Starting from $\eta
_{0}=n_{0}$, then $n_{1}$ is replaced by each of the subsequent $\eta _{i}$s.

\bigskip

\subparagraph{\textbf{Algorithm X: Samples }$(X_{1},\ldots ,X_{N_{X}+\ell })$
\textbf{given} $a\in (0,1)$, $\overline{\protect\sigma }>0$, $\ell \geq 0$}

\label{alg:N_X}

Step 1: $X\gets [\,]$, $\eta\gets n_0$

Step 2: $X \gets$ sample $(X_k)_{k=1,\ldots,\eta}$ under $P$

Step 3: Repeat

Step 4: \qquad segment $\leftarrow \text{ \textsc{SampleSingleRecord}}%
(a,n_{0},\eta )$

Step 5: \qquad If {segment is not `degenerate'}

Step 6: \qquad \qquad $X \gets [X, \text{segment}]$

Step 7: \qquad \qquad $\eta \gets \text{length}(X)$

Step 8: \qquad EndIf

Step 9: Until {segment is `degenerate'}

Step 10: If {$\ell>0$}

Step 11: \qquad $X\leftarrow \lbrack X,\text{\textsc{SampleWithoutRecordX}}%
(\eta ,\ell )]$

Step 12: EndIf

\subsection{Algorithm to Sample $X_{1},...,X_{N},N$ \label{Sec_Sample_M_all}}

The final algorithm for sampling $M,X_{1},...,X_{N},N$ is given next.

\subparagraph{\textbf{Algorithm M: Samples } $M,X_{1},...,X_{N},N$ \textbf{%
given} $a\in (0,1)$, $\protect\gamma <E\left( A_{1}\right) $, and $\overline{%
\protect\sigma }$}

\label{alg:final} ~\newline

Step 1: Sample $A_1,\ldots, A_{N_A}$ using Steps 1--9 from \textbf{Algorithm
S} with $S_n=\gamma n-A_n$.

Step 2: Sample $X_1,\ldots, X_{N_X}$ using Steps 1--9 from \textbf{Algorithm
X}.

Step 3: Calculate $N_a$ with (\ref{eq:defNmu}) and set $N=\max(N_A,N_X,N_a)$.

Step 4: If {$N>N_A$}

Step 5: \qquad Sample $A_{N_A+1},\ldots, A_N$ as in Step 10--12 from \textbf{%
Algorithm S} with $S_n=\gamma n-A_n$.

Step 6: EndIf

Step 7: If {$N>N_X$}

Step 8: \qquad Sample $X_{N_X+1},\ldots, X_N$ as in Step 10--12 from \textbf{%
Algorithm X}.

Step 9: EndIf

Step 10: Return $M(t_{i})=\max_{1\leq n\leq N}\left\{ -\log
A_{n}+X_{n}(t_{i})+\mu (t_{i})\right\} $ for $i=1,\ldots ,d$, and $%
X_{1},...,X_{N},N$.

\bigskip

%%%%%%%%%%%%%%%%%%%%%%%%%%%%%%%%%%%%%%%%%%%%%%%%%%%%%%%%%%%%%%%%%%%%%%%%%%%%%%%%%%%%%%%%%%%
%%% The acknowledgements
%\begin{acknowledgement}
%To include acknowledgements use the \verb|acknowledgement| environment.
%\end{acknowledgement}

%%%%%%%%%%%%%%%%%%%%%%%%%%%%%%%%%%%%%%%%%%%%%%%%%%%%%%%%%%%%%%%%%%%%%%%%%%%%%%%%%%%%%%%%%%%
%%% The bibliography
%
% BibTeX users please use
\bibliographystyle{spmpsci}
\bibliography{mybibfile}
% and then copy paste the contents of the .bbl file here for the final version.
%
% E.g.:

\end{document}